# Autonomic Cloud Computing: Open Challenges and Architectural Elements

Rajkumar Buyya[1,2], Rodrigo N. Calheiros[1], and Xiaorong Li[3]

[1]Cloud Computing and Distributed Systems (CLOUDS) Laboratory
Department of Computing and Information Systems
The University of Melbourne, Australia
E-mail: {rbuyya, rnc}@unimelb.edu.au

[2]Manjrasoft Pty Ltd, Melbourne, Australia

[3]Institute of High Performance Computing
A*STAR Institute, Singapore
E-mail: lixr@ihpc.a-star.edu.sg

*Abstract*—As Clouds are complex, large-scale, and heterogeneous distributed systems, management of their resources is a challenging task. They need automated and integrated intelligent strategies for provisioning of resources to offer services that are secure, reliable, and cost-efficient. Hence, effective management of services becomes fundamental in software platforms that constitute the fabric of computing Clouds. In this direction, this paper identifies open issues in autonomic resource provisioning and presents innovative management techniques for supporting SaaS applications hosted on Clouds. We present a conceptual architecture and early results evidencing the benefits of autonomic management of Clouds.

## I. Introduction

Cloud computing *"refers to both the applications delivered as services over the Internet, and the hardware and system software in the data centres that provide those services"*, according to Armbrust *et al.*[1], and *"is a utility-oriented distributed computing system consisting of a collection of inter-connected and virtualized computers that are dynamically provisioned and presented as one or more unified computing resource(s) based on service-level agreements established through negotiation between the service provider and consumers"* according to Buyya *et al.*[2]. Both definitions capture the real essence of this new trend in distributed systems, where both software applications and computing infrastructure are moved from private environments to third party data centres, and made accessible through the Internet. Cloud computing delivers infrastructure, platform, and software (applications) as subscription-based services in a pay-as-you-go model. In industry, these services are referred to as Infrastructure as a Service (IaaS), Platform as a Service (PaaS), and Software as a Service (SaaS), respectively.

To support end-user applications, service providers such as Amazon [3], HP [4], and IBM [5] have deployed Cloud data centers worldwide. These applications range from generic text processing software to online healthcare. Once applications are hosted on Cloud platforms, users are able to access them from anywhere at any time, with any networked device, from desktops to smartphones. The Cloud system taps into the processing power of virtualized computers on the back end, thus significantly speeding up the application for the users, who pay for the actually used services. However, management of large-scale and elastic Cloud infrastructure offering reliable, secure, and cost-efficient services is a challenging task. It requires co-optimization at multiple layers (infrastructure, platform, and application) exhibiting autonomic properties. Some key open challenges are:

- *Quality of Service (QoS).* Cloud service providers (CSPs) need to ensure that sufficient amount of resources are provisioned to ensure that QoS requirements of Cloud service consumers (CSCs) such as deadline, response time, and budget constraints are met. These QoS requirements form the basis for SLAs (Service Level Agreements) and any violation will lead to penalty. Therefore, CSPs need to ensure that these violations are avoided or minimized by dynamically provisioning the right amount of resources in a timely manner.

- *Energy efficiency.* It includes having efficient usage of energy in the infrastructure, avoiding utilization of more

resources than actually required by the application, and minimizing the carbon footprint of the Cloud application.

- *Security*. Achieving security features such as confidentiality (protecting data from unauthorized access), availability (avoid malicious users making the application unavailable to legitimate users), and reliability against Denial of Service (DoS) attacks. The DoS is critical because, in a dynamic resource provisioning scenario, increase in the number of users causes automatic increase in the resources allocated to the application. If a coordinated attack is launched against the SaaS provider, the sudden increase in traffic might be wrongly assumed to be legitimate requests and resources would be scaled up to handle them. This would result in an increase in the cost of running the application (because provider will be charged by these extra resources) as well as a waste of energy.

As Clouds are complex, large-scale, and heterogeneous distributed systems (e.g., consisting of multiple Data Centers, each containing 1000s of servers and peta-bytes of storage capacity), management is a crucial feature, which needs to be automated and integrated with intelligent strategies for dynamic provisioning of resources in an autonomic manner. Effective management of services becomes fundamental in platforms that constitute the fabric of computing Clouds; and to serve this purpose, autonomic models for PaaS (Platform as a Service) software systems are essential.

Autonomic systems exhibit the ability of self-monitoring, self-repairing, and self-optimizing by constantly sensing themselves and tuning their performance [6]. Such autonomic features are also exhibited by market economy, where resources/services are priced so as to maintain equilibrium in the supply and demand. Clouds constitute an interesting venue to explore the use of autonomic features, because of their dynamism, large scale, and complexity.

In this direction, this paper presents our early steps towards innovative autonomic resource provisioning and management techniques for supporting SaaS applications hosted on Clouds. Steps towards this goal include (i) development of an autonomic management system and algorithms for dynamic provisioning of resources based on users' QoS requirements to maximize efficiency while minimizing the cost of services for users and (ii) creation of secure mechanisms to ensure that the resource provisioning system is able to allocate resources only for requests from legitimate users. We present a conceptual model able to achieve the aforementioned goals and present initial results that evidence the advantages of autonomic management of Cloud infrastructures.

II. RELEVANT WORK

Autonomic management [6], [25] is a desired feature for any large scale distributed system and even more important in dynamic infrastructures such as Clouds. Autonomic systems are self-regulating, self-healing, self-protecting, and self-improving. In other words, they are self-managing. Initial investigation on developing autonomic based systems in both academia and industry has been already carried out. Parashar and Hariri [11] reported an overview of the early efforts in developing autonomic systems for storage management (OceanStore [7], Storage Tank [8]), computing resources (Oceano [9]), and databases (SMART DB2 [10]). Computing Grids have benefited from the application of autonomic models for management of resources and the scheduling of applications [11], [12], [13], [14]. Even though none of these platforms considers energy-efficiency as a high-priority parameter to be optimized, the success in autonomic management for Grid applications demonstrates potential of integrating autonomic models in Cloud Computing.

CometCloud [15] implements an infrastructure for autonomic management of workflow applications on Clouds. Recently other works [16],[17],[18] explored provisioning of resources for Grid and Cloud applications. However, they do not support an integrated solution for security-enforced, cost-effective, energy efficient, and dynamic resource provisioning, which are key open issues.

Solutions for secure Cloud platforms have been proposed in the literature [19]. However, existing works are yet to address issues related to recognition of attacks against SaaS with the aim of exploiting elasticity. A step towards this goal has been given by Sqalli et al. [20]. Their EDoS-Shield system is able to detect and mitigate distributed denial of service attacks against Clouds. However, research is required to determine if the same or similar techniques can be applied for thwarting attacks against elastic infrastructures.

Amazon Elastic MapReduce has enabled its customers to dynamically modify the size of their running job flows. Using their API, customers have the flexibility to add or remove nodes based on the changing capacity needs of their job flow. However, this service does not offer automatic provisioning of new nodes based on end-user demands/QoS.

III. ARCHITECTURE FOR AUTONOMIC CLOUD MANAGEMENT

As we aim towards the development of autonomic resource provisioning and management techniques for supporting SaaS applications hosted on Clouds, the following aspects were identified as essential:

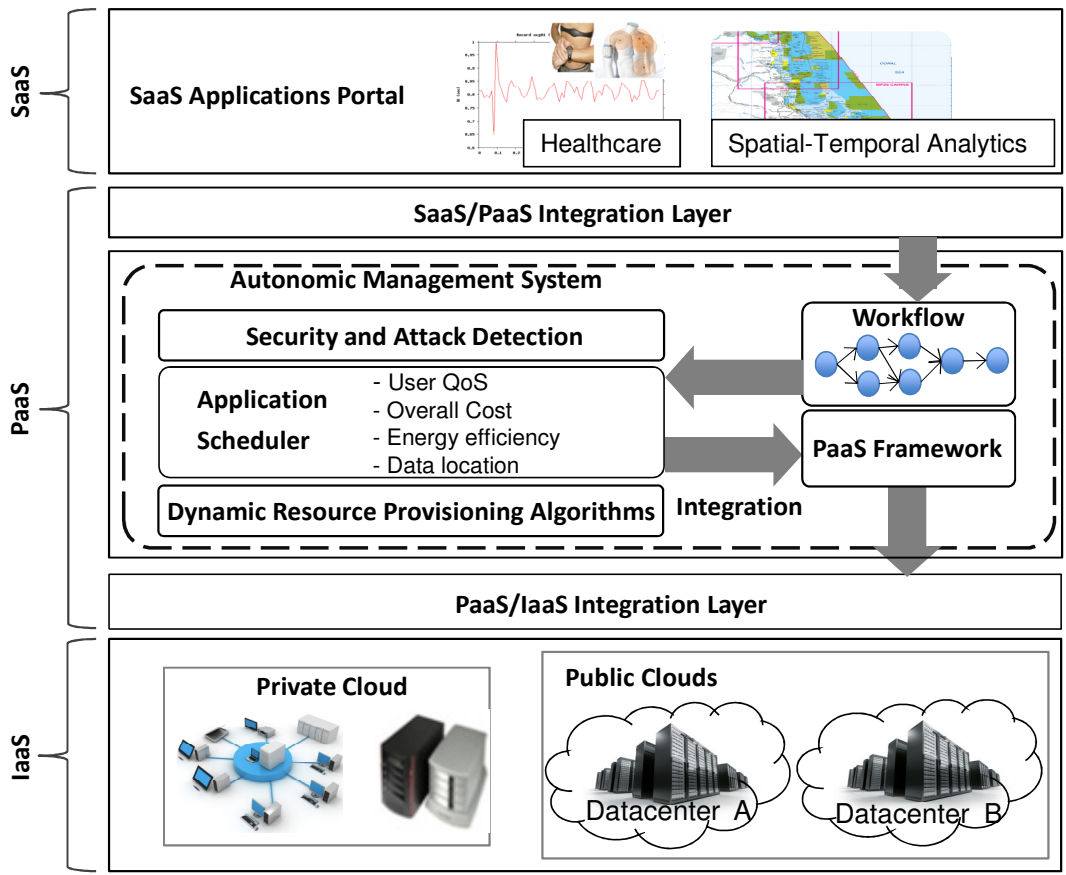

Figure 1. System architecture for autonomic Cloud management.

- Development of an autonomic management system and algorithms for dynamic provisioning of resources based on users QoS requirements to maximize efficiency while minimizing the cost of services for users.
- Creation of secure mechanisms to ensure that the resource provisioning system is able to allocate resources only for requests from legitimate users.

Figure 1 shows the high-level architecture enabling autonomic management of SaaS applications on Clouds. The main components of the architecture are:

- *SaaS Application Portal*: This component hosts the SaaS application using a Web Service-enabled portal system. Users or brokers acting on their behalf submit service requests from anywhere in the world to these SaaS applications.
- *Autonomic Management System and PaaS Framework*: This layer serves as a Platform as a Service. Its architecture comprises of autonomic management components to be integrated in the PaaS level, along with modules enforcing security and energy efficiency. User QoS-based application scheduler and dynamic resource provisioning algorithms are added as plug-ins.
- *Infrastructure as a Service*: This layer comprises distributed resources provided by private (enterprise networks) and public Clouds. Enterprise networks could leverage the resources in public Clouds by leasing them according to their user requirements, as and when needed.

SaaS is described as a software application deployed as a hosted service and accessed over the Internet. This model provides a scalable way for service providers and ISVs (Independent Software Vendors) to deliver their existing and/or new software applications to end-users without having to worry about the expertise or the capital budget to purchase, install, and manage large IT infrastructure. In order to manage the SaaS applications in large scale, the PaaS layer has to coordinate the Cloud resources according to the SaaS requirements, which is ultimately the user QoS. This coordination requires the PaaS layer to handle the scheduling of applications and resource provisioning such that the user

QoS is satisfied and also it does not make the provisioning too costly to the PaaS service provider.

The autonomic management system incorporates the following services in the PaaS layer: Security and attack detection, application scheduling, and dynamic provisioning. The autonomic manager is composed by the following components, with specific roles:

- *Application Scheduler.* The scheduler is responsible for assigning each task in an application to resources for execution based on user QoS parameters and the overall cost for the service provider. This scheduler is aware of different types of applications such as independent batch applications (such as Bag of Tasks), web multi-tier applications, and scientific workflows (where tasks have dependencies that have to be managed) executed in Clouds. Depending on the computation and data requirements of each application, it directs the dynamic resource-provisioning component to instantiate or terminates specified number of compute, storage, and network resources while maintaining a queue of tasks to be scheduled. Execution of the application also may require data transfer between Clouds, which is also handled by this component. This logic is embedded as multi-objective application scheduling algorithms [21]. This heuristic-based algorithm focuses on QoS parameters such as response time, cost of service usage, energy consumption, maximum number of resources available per unit price, and penalties for service degradation.

- *Energy-efficient scheduler.* One of the main objectives to be optimized during the application scheduling process is energy utilization. Applications need to be scheduled in resources in such a way that their total energy consumption is minimized. However, the algorithm has to achieve this goal without compromising SLAs and cost. This is a multi-objective optimization problem with conflicting goals. An aspect of this problem that makes it even more challenging is the fact that energy consumption holds a non-linear relationship with cost and performance. Search for a solution for such a challenging and relevant problem is one of the main challenges of this research.

- *Dynamic Resource Provisioning Algorithms.* This component implements the logic for provisioning and managing virtualized resources in private and public Cloud environments based on the resource requirements as directed by the application scheduler. This is achieved by dynamic negotiation with Cloud IaaS providers for the right type of resource for a certain time and cost by taking into account the past execution history of applications and budget availability. The resource-provisioning module is complemented with prediction-based algorithms that rely on market-oriented provisioning practices, for handling any change in spot prices. In particular, these algorithms perform the following tasks:

    o Dynamic resource allocation: Scaling in/out (expanding/shrinking of resources) will be carried out using an online instantiation mechanism where compute, storage and network services will be leased on the fly. Resources are terminated once they are no longer needed by the system.

    o Prediction for resource selection: As the cost of using resources depends on the duration and type of resources provisioned, a prediction mechanism will be implemented that takes into account historic execution statistics of SaaS applications. Based on prediction of time and cost, this component will control the resource plug-in component to allocate either the spot-instances or the fixed price instances of IaaS resources. We also plan to conduct resource-pricing design based on these predictions. The prediction will be based on the supply and demand for resources, similar to market-oriented principles used for reaching equilibrium state [2].

- *Security and Attack Detection*: This component implements all the checks to be performed when requests are received in order to evaluate their legitimacy. This prevents the scaling-up of resources to respond to requests created with the intention of causing a Denial of Service or other forms of cyber-attacks. The module must be able to distinguish between authorized access and attacks, and in case of suspicion of attack, it can either decide to drop the request or avoid excessive provision of resources to it. To achieve it, techniques already in use for detection of DDoS attacks need to be adapted to be able to handle exclusive characteristics of Cloud systems. In this sense, this module has to work as a "DDoS Detection as a Service" for the PaaS middleware.

IV. DATA ANALYTICS WORKFLOW ENGINE: A CASE STUDY

In order to demonstrate the importance and the impact of autonomic Cloud management, we present in this section a case study of autonomic Cloud management in the context of workflow applications for spatial-temporal data analytics for online prediction of dengue fever outbreaks in Singapore and their deployment on Clouds.

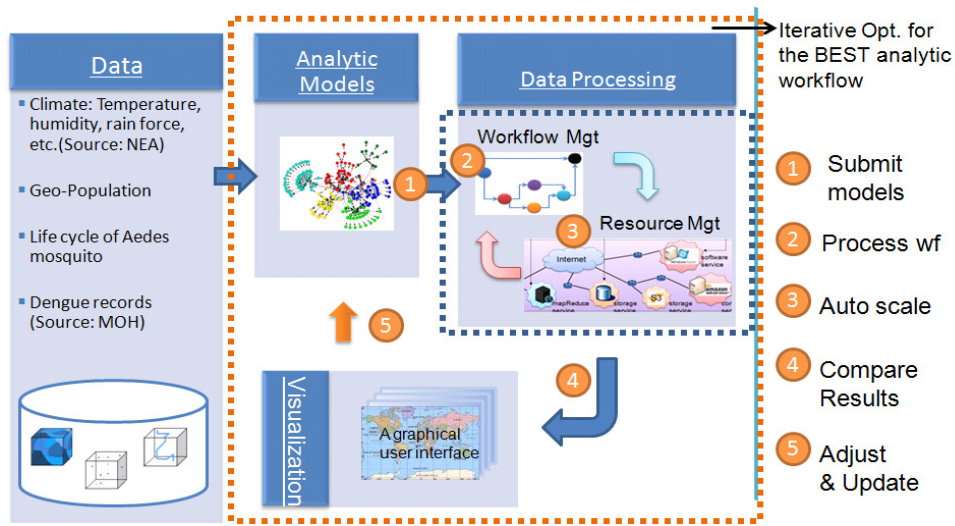

Figure 2. Flows of workflow-enabled scalable spatial-temporal analysis.

Dengue is a mosquito-borne infectious disease that occurs especially in tropical regions such as South America and Southeast Asia. According to the World Health Organization (WHO), there are 2.5 billion people in the world living in dengue endemic places, which makes it a major international public health concern. This is further aggravated in densely populated regions, where the disease can spread quickly. Therefore, prediction and control of dengue is a very important public health issue for Singapore [22], and this motivated the development of prediction models for dissemination of the disease in the country.

The application data requirement comprises multi-dimensional data containing information such as reported dengue incidents, weather parameters, and geographic information. Incidence data can reach hundreds of MB, and the associated weather data can easily take up a few GBs. For example, at a daily resolution, a single output variable from the ECHAM5 climate model comprises 300,000 spatial points multiplied by 365,250 temporal intervals per century per scenario. Application users must be able to trace the number of dengue incidences by day, week, month, and year from 1960s to 2011.

The processing time required to extract the data, model it, and interpolate for visualization is about 30 minutes in total for processing 1-day data set on a workstation with an Intel dual core 2.93GHz CPU and 4GB of memory. Moreover, in order to be of practical value in the case of dengue outbreak, the system must be able to dynamically allocate resources and optimize the application performance on Cloud infrastructures (private, public, or hybrid Clouds) to reduce the processing time and enable real-time spatial and temporal analysis with shorter turnaround time.

From the above, we can clearly notice that autonomic Cloud technologies are paramount for the goals of timely prediction of dengue dissemination, so that health agencies can be mobilized to react to the incident. We now describe our Cloud-enabled Workflow Engine used in this case study.

*A. Cloud Workflow Engine and Autonomic Management*

The Cloudbus Workflow engine [24] is an extension of a Grid-based workflow management system [23] supporting the execution of workflows in private, public, and hybrid Clouds. Initially, it supported features such as GUI-based description of workflows, application composition, data movement across Clouds, and task scheduling and management. It has been further extended to incorporate autonomic management capabilities based on iterative optimizations.

The overview of the autonomic workflow management system and its use in data analytics application is depicted in Figure 2. The performance requirements are achieved by partition of the data in different parallel tracks and execution of such tracks on multiple virtual machines simultaneously. To achieve this, the system autonomically optimizes its performance and finds the optimal provisioning for utilization and performance optimization.

The iterative optimization is designed for workflow analytical applications in which a subset of the analytic tasks/functions is repeated during the analytics, forming a sort of "loop" in the workflow execution. When such loops are detected in the applications, the workflow engine profiles the early execution of tasks, storing information about their execution time. This profile information is used for optimal provisioning purposes in terms of *cost* and *execution time (makespan)*. Hence, the performance of running the data

**Step 0.** Initiate the Cloud resources to execute the tasks.

**Step 1.** Apply a greedy algorithm to minimize the makespan ignoring cost and resource constraints.

**Step 2.** Apply an initial schedule that fully utilize the allocated machines by scheduling extra tasks to resources as long as it does not increase the makespan.

**Step 3.** Analyze whether downgrading the public Cloud instance type still enable completion of the workflow within the same time slot. If so, utilize the simpler and cheaper instance type.

**Step 4.** Run the tasks on the schedule nodes.

Figure 3. Algorithm for iterative optimization.

analytics program is continuously improved by the system, which autonomically scales up and down provisioned resources to meet the users' performance requirements.

For the purposes of performing the dynamic provisioning, the optimization problem solved by the scheduler consists of an initial schedule $S$ that allocates all the workflow tasks from the workflow graph G to Cloud resources considering precedence constraints. We define Time $t(S)$ and Cost $m(S)$ as the completion time and monetary cost of the schedule S, respective. The iterative optimization technique aims to derive an optimal schedule $S_{opt}$ to achieve $t_{min}(SG)$ and $m_{min}(SG)$. As the problem of mapping workflow tasks onto distributed heterogeneous resources to achieve multi-objective optimization is NP-complete, we proposed a heuristic algorithm to achieve sub-optimal solution and improve the results iteratively during the workflow execution. The algorithm is described in Figure 3.

The autonomic adaptive workflow engine design allows the system to select the most suitable resources according to the user requirements (e.g., update frequency, cost, etc), schedule the privacy-sensitive data in private resources, and tolerate faults when failure happens. Provisioning of Cloud resources and scheduling of workflow tasks are automatically performed based on a budget constraint, and the system schedules tasks to resources that can optimize the performance in terms of the total execution time while satisfying eventual budget requirements for application execution.

Finally, it is worth noting that autonomic execution of workflows for dengue fever prediction is just one of the possible scenarios for application of autonomic management of Clouds. As Cloud platforms become more widespread as the infrastructure of choice for several domains such as e-health, e-Science, e-government, and e-commerce, the need for autonomic management of Clouds will spread across all these domains. Nevertheless, the general principles and architectural aspects of such autonomic platforms will follow architectural elements presented in this paper, regardless the application domain.

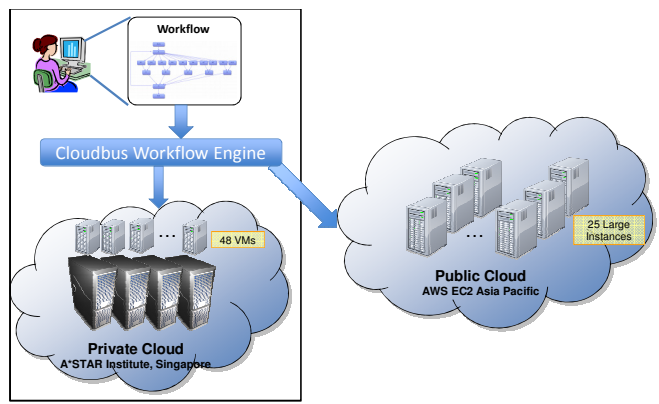

Figure 4. Experimental testbed.

## V. PERFORMANCE EVALUATION

We present an evaluation of the autonomic iterative optimization feature of the workflow engine. The experimental testbed, depicted in Figure 4, consists of a hybrid Cloud composed of a local infrastructure (located in an A*STAR Institute, Singapore) containing four nodes, each of which had 24 cores (hyper-threaded) 2.93 GHz processor and 96 GB of memory and running 48 Linux CentOS 5.8 virtual machines with 2 or 4 cores and 4 GB of memory. This local infrastructure is complemented by 25 Amazon EC2 large compute instances (2 cores with 2 ECU and 7.5 GB of memory) deployed in the region of Asia Pacific (South East). The application utilized for the experiment is the dengue fever prediction application, which utilizes historical dengue cases and climate data from 2001 to 2010.

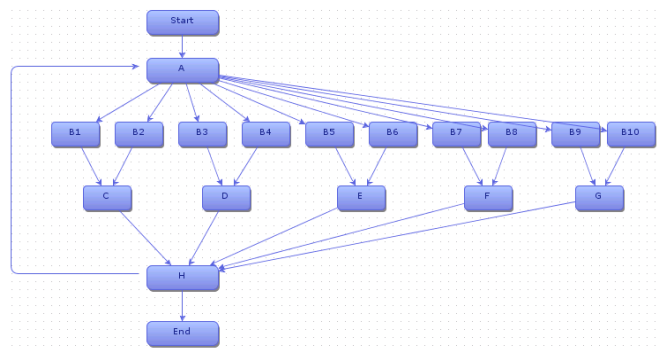

Figure 5. Iterative workflow model of the dengue fever prediction model used in the experiments. The iteration happens between tasks H and A, as depicted in the figure.

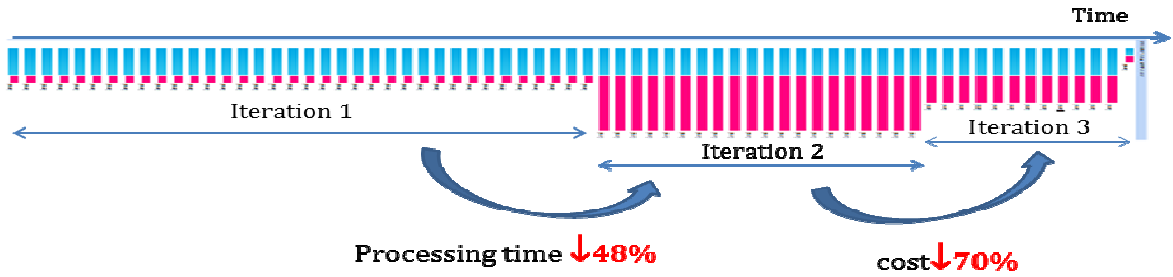

Figure 6. Effect of the iteration optimization of workflow execution in the dengue fever prediction model.

The predictive model utilized is based on a sliding window, where parameter variables are adjusted periodically through the comparison of real dengue cases with the prediction results. Such interactive analytics model can be mapped to the workflow in Figure 5.

The iterative scheduling algorithm searches the suboptimal solutions aggressively by using information of previous iterations of the workflow execution. The iteration loop occurs between tasks labeled as H and A as shown in Figure 5. As a consequence of the iterative loop, tasks labeled from B to G are re-executed as each new iteration starts, with information related to a different time window being used as tasks input.

As each of the iterations completes, the workflow system computes the expected execution time of tasks and the cost of keeping the current amount of resources for execution. If changes in the number of available resources can lead to substantial improvement in either makespan or cost, the number of provisioned resources is scaled up or down. This enables the system to fine-tune and adapt the provisioning and scheduling according to the characteristics of the workflow tasks and the execution environments.

Figure 6 presents the results of variation of number of resources provisioned by the workflow engine in different iterations of the execution of the prediction model. After collecting information about the actual execution time of the tasks at the first iteration, the number of provisioned resources was corrected so that the tasks were consolidated in fewer Cloud resources. Further corrections where applied between iterations 2 and 3. Overall, the autonomic iterative optimization feature of the workflow engine enabled a reduction of execution time of 48% and reduction of cost of public Cloud utilization in 70% compared to a greedy solution for provisioning and scheduling of workflow applications in Clouds.

## VI. CONCLUSIONS AND FUTURE WORK

The growing adoption of Cloud computing as the preferred solution for hosting business and academic systems evidences the need for better solutions for management of such platforms. Considering that Cloud platforms are typically composed of thousands of physical hosts and virtual machines, connected by many network elements, management of such infrastructures is also becoming a challenging task. Furthermore, as Clouds get bigger visibility as a strategic asset for organizations, they will also increasingly become the target of cyber-attacks.

This paper presented our first steps towards an autonomic Cloud platform able to handle several of the above problems. Such a platform will be able to dynamically provision Cloud resources to applications in such a way that Quality of Service expectations of users are met with an amount of resources that optimizes the energy consumption required to run the application. Moreover, the platform will also be able to differentiate regular requests from DDoS attacks against the infrastructure, avoiding the wastage of energy and budget caused by provision of resources to illegitimate requests.

Our early experiments demonstrate the potential of the platform to optimize workflow applications, which are complex applications where dependencies between tasks exist and have to be respected by the platform.

As future work, we will implement more dynamic provisioning algorithms that are QoS and security-aware and energy efficient, and will demonstrate their effectiveness with real applications from domains like disaster management, environment data analysis and healthcare, as we identify these as target areas that can benefit the most from an autonomic Cloud system.

ACKNOWLEDGEMENTS

This work is partially supported by the Australian Research Council Future Fellowship grant. We would like to thank our colleagues Jia Yu, Suraj Pandey, Sifei Lu, Long Wang, Henry Palit, and Qin Zheng for their contribution towards Workflow Engine. We thank Deepak Poola and Nikolay Grozev for their comments on improving the paper.